\documentclass[twocolumn,doublespacing,showpacs,preprintnumbers,amsmath,amssymb]{revtex4}
\usepackage{amssymb}
\usepackage{txfonts}
\usepackage{graphicx}
\usepackage{subfigure}
\usepackage{dcolumn}
\usepackage{bm}
\usepackage{sidecap}

\begin{document}
\title{The Security of SARG04 Protocol in Plug and Play Quantum Key Distribution system with an Untrusted Source}
\author{Bingjie Xu}
\author{Xiang Peng}
\thanks{xiangpeng@pku.edu.cn.}
\author{Hong Guo}
\thanks{hongguo@pku.edu.cn.}
\affiliation{CREAM Group, State Key Laboratory of Advanced Optical Communication Systems and Networks (Peking University) and Institute of Quantum Electronics, School of Electronics Engineering and Computer Science, Peking University, Beijing 100871,
PR China}
\date{\today}
\begin{abstract}
The SARG04 protocol is one of the most frequently used protocol in commercial plug-and-play quantum key distribution (QKD) system, where an eavesdropper can completely control or change the photon number statistics of the QKD source. To ensure the security of SARG04 protocol in plug-and-play QKD system with an unknown and untrusted source, the bounds of a few statistical parameters of the source need to be monitored.  An active or a passive source monitor schemes are proposed to verify these parameters. Furthermore, the practical issues due to statistical fluctuation and detection noise in the source monitoring process are quantitatively analyzed. Our simulation results show that the  passive scheme can be efficiently applied to plug-and-play system with SARG04 protocol.
\end{abstract}
\pacs{03.67.Dd, 03.67.Hk}
\maketitle

\section{Introduction}
Quantum key distribution (QKD) provides a means of sharing a secret key between two parties (Alice and Bob) in the presence of an eavesdropper (Eve).
The single-photon (e.g. BB84~\cite{BB84} and SARG04~\cite{SARG04}), entanglement-based (e.g. E91~\cite{E91}) and continuous variable (e.g. GG02 \cite{GG02}) QKD protocols have proved to be unconditionally secure under ideal (source, channel, detection and postprocessing) assumptions \cite{Shor_Preskill_00, ILM_07,GLLP_04,SARG_06,DIQKD_07,CVQKD1,CVQKD2,RMP_09}. In practical QKD systems, the security assumptions are not completely satisfied and security loopholes exist \cite{Rev_Lo_08}. Real implementations of QKD may deviate from the ideal models in security proofs, such as laser with intensity fluctuation \cite{{UntruQKD_Wang_PRA_07},UntruQKD_Wang_APL_07}, detectors with mismatched detection efficiency \cite{Eff_Mismatch_Makarov_PRA_06, Time_Shift_Att_07,Time_Shift_Att_08,Eff_Mismatch_Makarov_QIC_08,Eff_Mismatch_Lydersen_QIC_10,{Time_Shift_Att_09}}, or detection blinding~\cite{Det_Blind_09,Det_Blind_NatPho_10, Det_Blind_Exp_10}. The unconditional security of practical QKD systems will be compromised if these loopholes are not included in general security analysis or no counter measures are made. For instance, the ideal security proof for the BB84 protocol was given when a single-photon source was assumed \cite{Shor_Preskill_00}, while highly attenuated laser source is often used in real experiment, where the source sometimes produces multi-photon states. Due to the channel loss and these multi-photon states, Eve can perform the photon-number-splitting (PNS) attack~\cite{PNS_00}. Lately, more general security analysis for the BB84 protocol with weak coherent laser source and semi-realistic models were given \cite{{ILM_07},GLLP_04}. Furthermore, several methods (such as decoy state \cite{Decoy_Hwang_03,Decoy_Lo_04, Decoy_Lo_05,Decoy_Ma_PRA_05, Decoy_Wang_05,Decoy_Wang_PRA_05,Decoy_Passive_07} and SARG04 \cite{SARG04} protocols) have been proposed to fight against the multi-photon loophole.

The security loophole considered in this paper is the untrusted source problem \cite{TrojanAttack_06,UntruQKD_Wang_07,UntruQKD_Zhao_08,UntruQKD_Zhao_09,UntruQKD_Peng_08,UntruQKD_Peng_09,UntruQKD_Wang_08,UntruQKD_Wang_09,UntruQKD_Guo_09,UntruQKD_Wang_10,Untrusted_Xu_10}.
In the standard security analysis of some protocols (such as BB84, decoy state, and SARG04 protocols), the photon number distribution (PND) of the QKD source is assumed to be fixed and known to Alice and Bob, which is defined as a trusted  source. However, in a one-way QKD system, the intensity   fluctuation from the laser source and the parameter fluctuation from the optical devices cause the assumption of the trusted source to fail \cite{UntruQKD_Wang_PRA_07,UntruQKD_Wang_APL_07,UntruQKD_Guo_09}. More seriously in a two-way plug-and-play QKD system, Eve can even control or change the PND of the QKD source in principle, such that the source is unknown and untrusted \cite{UntruQKD_Zhao_08}. To solve the untrusted source problem, the statistical characteristics of the QKD source need to be monitored in real experiment~\cite{UntruQKD_Peng_08}. Many theoretical researches have been done on the security analysis for BB84 and decoy state protocols with an untrusted source~\cite{UntruQKD_Zhao_08,UntruQKD_Zhao_09,
UntruQKD_Peng_09,UntruQKD_Wang_08,UntruQKD_Wang_09,UntruQKD_Wang_10,Untrusted_Xu_10}, and the real-time source monitoring for both one-way and two-way systems have been demonstrated experimentally~\cite{UntruQKD_Peng_08,UntruQKD_Zhao_09,UntruQKD_Guo_09}.

As is pointed out in \cite{RMP_09}, the SARG04 protocol is more robust than BB84 against the PNS attack, and has been applied in commercial plug-and-play QKD system~\cite{id}. However, this protocol also suffer from the untrusted-source problem. In this paper, rigorous security
analysis for the SARG04 protocol with an untrusted source is given, and the lower bound of secure key rate is devised if the ranges of a few statistical parameters of the untrusted source are known. Then, an active and a passive schemes are proposed to monitor these parameters. Furthermore, the practical issues of finite data size and detection noise are quantitatively analyzed.

\section{Security analysis for the SARG04 Protocol with an untrusted source}

The security key rate of the SARG04 protocol is \cite{SARG_06}
\begin{eqnarray} \label{Eq:R_SARG}
\nonumber {R} &=&- {Q_\mu }f({E_\mu }){H_2}({E_\mu }) + {Q_1}[1 - H_2({Z_1}|{X_1})] \\
& &+{Q_2}[1 - H_2({Z_2}|{X_2})],
\end{eqnarray}
where $Q_\mu $ and $E_\mu $ are the total count rate and quantum bit error rate (QBER) respectively, $ Q_{1(2)} $ is the gain of the 1(2)-photon state, $ Z_{1(2)} $ and $ X_{1(2)}$ are random variables characterizing the phase and bit errors for the 1(2)-photon state respectively, $f(x)$ is the error correction efficiency, and $ H_2(x)=-x\log_2(x)-(1-x)\log_2(1-x)$ is the Shannon entropy function. Suppose $p_{X1(2)}$ denote the probability that bit flip without phase flip occurs on 1(2)-photon state, $p_{Z1(2)}$ denote the probability that phase flip without bit flip occurs on 1(2)-photon state, and $p_{Y1(2)}$ denote the probability that both bit flip and phase flip occur on 1(2)-photon state. Let $e_i$ $(e_{pi})$ denote the bit (phase) error rate for $i$-photon state, and $e_{1(2)}=p_{X1(2)}+p_{Y1(2)},e_{p1(2)}=p_{Z1(2)}+p_{Y1(2)}$. It has been proved for one-way postprocessing that \cite{SARG_06},
\begin{eqnarray}\label{Eq:e1 e2}
\nonumber &{p_{X1}}&= {e_{1}} - a,\ {p_{Z1}} = \frac{3}{2}{e_{1}} - a,\ {p_{Y1}} = a,\\
&{p_{X2}}& = {e_{2}} - b,\ {p_{Z2}}\le x{e_{2}} + g(x) - b,\ {p_{Y2}} = b,
\end{eqnarray}
where $g(x)=[3-2x+(6-6\sqrt{2}x+4x^2)^{1/2}]/6$, and $e_{1}/2\le a\le e_{1}$, $0\le b\le e_{2}$. Based on Eq.~(\ref{Eq:e1 e2}), one has $H_2(Z_1|X_1)
\le H_2^{\max}(Z_1|X_1)$, where $H_2^{\max}(Z_1|X_1)=-H_2(e_1)-e_1\log_2(e_1^2)-(1-2e_1)\log_2(1-2e_1)$, and $H_2(Z_2|X_2)\le H_2(Z_2)\le H_2(e_{p2}^{opt})$, where $e_{p2}^{opt}=\mathop {\max }\limits_{x}\{ x{e_2} + g(x)\}$. Then,
\begin{eqnarray} \label{Eq:R_SARG1}
\nonumber {R}&\ge &- {Q_\mu }f({E_\mu }){H_2}({E_\mu })+ {Q_1}[1 - H_2^{\max}({Z_1}|{X_1})] \\
&&+ {Q_2}[1 - H_2(e^{opt}_{p2})].
\end{eqnarray}In order to calculate the final secure key rate, one needs a good estimation of $ Q_{1(2)} $ and $ e_{1(2)}$. There are a few methods to approach the target. One is proposed by GLLP \cite{GLLP_04}, where all the losses and errors are assumed from the 1-photon and 2-photon states, and $ Q_{1(2)} $ and $ e_{1(2)}$ are overestimated. Another is the decoy state method \cite{Decoy_Hwang_03,Decoy_Lo_04,Decoy_Lo_05,Decoy_Ma_PRA_05, Decoy_Wang_05, Decoy_Wang_PRA_05,Decoy_Passive_07}, which can accurately estimate the parameters. Here, we consider the SARG04 protocol combined with decoy state method.

A fundamental assumption in the decoy state protocol with a trusted source is $e_{n}=e_{n}^s =e_{n}^d$ and $Y_{n}=Y_{n}^s = Y_{n}^d$ \cite{Decoy_Hwang_03}, where $Y_{n}$ is the yield of $n$-photon state and the superscript $s(d)$ means the signal (decoy) source. The optimal estimation, with applying infinite decoy states, converges to~\cite{SARG_06},
\begin{eqnarray}\label{Eq:Trusted source}
\nonumber {Y_{n}} &=& {\eta _n}(\frac{{{e_{\det}}}}{2} + \frac{1}{4}) + \frac{1}{2}(1 - {\eta _n}){Y_{0}},\\
{e_{n}} &=& [{\eta _n}\frac{{{e_{\det}}}}{2} + \frac{1}{4}(1 - {\eta _n}){Y_{0}}]/{Y_{n}},
\end{eqnarray}
where $\eta_n$ is the probability for $n$-photon state to arrive at Bob's detector, $Y_0$ is the dark count rate of Bob's detector, and $e_{\det}$ is the probability that a photon hit the erroneous detector in Bob's side. Then, one has $Q_{1(2)}=P_{1(2)}Y_{1(2)}$, where $P_{1(2)}$ is the probability for Alice to send out 1(2)-photon state that is fixed and known to Alice and Bob with a trusted source.

In the untrusted source case, the assumptions of $e_{n}^s = e_{n}^d$ and $Y_{n}^s = Y_{n}^d$ are broken \cite{UntruQKD_Zhao_08,UntruQKD_Wang_07}, and the results in Eq.~(\ref{Eq:Trusted source}) no longer hold. One needs new methods to estimate $ Q_{1(2)} $, which has been an open question. Fortunately, the results in~\cite{UntruQKD_Zhao_08,UntruQKD_Wang_08} provide two new ways to estimate the lower bound of $Q_1$ for BB84 protocol combined with 3-intensity decoy state methods. However, in SARG04 protocol, both 1-photon and 2-photon states have positive contributions to the secure key rate. Thus, the main task for the SARG04 protocol with an untrusted source is to derive the lower bound of $Q_2$. We find a modification of the method in \cite{UntruQKD_Wang_08} will approach this task. In the following, the lower bound of $Q_2$ is calculated for the SARG04 protocol combined with 4-intensity decoy state method in untrusted source scheme.

In a SARG04 protocol combined with 4-intensity decoy state method, Alice randomly sends four kinds of sources: vacuum, decoy-1, decoy-2 and signal source, with probability $p_0$, $p_1$, $p_2$, and $p'$, respectively. In the trusted source scheme, the source is controlled completely by Alice, and the quantum states of vacuum, decoy-1, decoy-2 and signal sources are expected to be $\rho_{0}=\left| 0 \right\rangle\left\langle 0 \right|$, $\rho_{1}=\sum_{n = 0}^\infty {a_n \left| n \right\rangle }\left\langle n \right|$, $\rho_{2}=\sum_{n = 0}^\infty {b_n \left| n \right\rangle }\left\langle n \right|$ and $\rho _s = \sum_{n = 0}^\infty {a'_n \left| n \right\rangle } \left\langle n \right|$, respectively, where $\{a'_n,\ a_n,\ b_n\}$ are fixed and known. In the untrusted source scheme, the source is controlled and prepared by Eve (as shown in Fig.~\ref{fig: Source Monitor}(a)), and $\{a'_n,\ a_n,\ b_n\}$ are unknown.

Suppose Alice sends $M$ optical pulses to Bob totally. In a real experiment, one could observe the following parameters: $N_s,$ $N_{d1(2)}$, and $N_0$ (the number of counts caused by signal, decoy-1(2), and vacuum sources, respectively). Then the count rates for signal, decoy-1(2), and vacuum sources are $Q_\mu={N_s}/{p'M}$, $Q_{d1(2)}={N_{d1(2)}}/{p_{1(2)}M}$, and $Y_0={N_0}/{p_0M}$, respectively. Denote the lower (upper) bound of $\{a'_n,\ a_n,\ b_n\}$ as $\{{a'_n}^{L(U)},\ a_n^{L(U)},\ b_n^{L(U)}\}$. One can rigorously prove that (see Appendix A for details)
\begin{equation}\label{Eq: Q1L}
{Q_1} \ge \frac{{a{'_2}^L{Q_\mu } - a_2^U{Q_{d1}} - (a{'_2}^La_0^U - a{'_0}^La_2^U){Y_0}}}{{a{'_2}^La_1^U - a{'_1}^La_2^U}}
\end{equation}
under condition
\begin{equation}\label{Eq:condition 0}
\frac{{{a'_k}^{L}}}{{a_k^{U}}} \ge \frac{{{a'_2}^{L}}}{{a_2^{U}}} \ge \frac{{{a'_1}^{L}}}{{a_1^{U}}},\ (\rm{for\ all}\ k \ge 3),
\end{equation}
and
\begin{widetext}
\begin{equation}\label{Eq: Q2L}
{Q_2} \ge \frac{{a{'_3}^L{Q_{d1}} - a_3^U{Q_\mu } - (a{'_3}^La_0^U - a{'_0}^La_3^U){Y_0} - (a{'_3}^La_1^U - a{'_1}^La_3^U)\frac{{{Q_{d2}} - b_0^L{Y_0}}}{{b_1^L}}}}{{c(a{'_3}^La_2^U - a{'_2}^La_3^U)}}
\end{equation}
\end{widetext}
under conditions
\begin{equation}\label{Eq:condition1}
\frac{{{a'_k}^{L}}}{{a_k^{U}}} \ge \frac{{{a'_3}^{L}}}{{a_3^{U}}} \ge \frac{{{a'_2}^{L}}}{{a_2^{U}}} \ge \frac{{{a'_1}^{L}}}{{a_1^{U}}},\ (\rm{for\ all}\ k \ge 4),
\end{equation}
and
\begin{equation}\label{Eq:condition2}
c=1+\frac{a_3^{U}{a'_1}^{L} - {a'_3}^{L}a_1^{U}}{{a'_3}^{L}a_2^{U} - a_3^{U}{a'_2}^{L}}\frac{b_2^L}{b_1^L}> 0.
\end{equation}

When one consider the contribution from only 1-photon state for the SARG04 protocol, the final secure key rate is
\begin{equation}\label{Eq:R 1photon}
{R_{1photon}} \ge - {Q_\mu }f({E_\mu }){H_2({E_\mu }) + {Q_1}[1 - H_2^{\max}}({Z_1}|{X_1})] .
\end{equation}
The parameters $\{$${a'_0}^L$, $a_0^U$, ${a'_1}^{L}$, $a_1^U$, ${a'_2}^{L}$, $a_2^U\}$ need to be verified to estimate the gain of 1-photon state in Eq.~(\ref{Eq: Q1L}), after which one has  $e_1\le E_\mu Q_\mu/{Q_1}.$ Then one can calculate the secure key rate as Eq.~(\ref{Eq:R 1photon}). This case is defined as {\bf Case-1}. When one consider the contributions from both 1-photon and 2-photon states, the parameters $\{$${a'_0}^L$, $b_{0}^L$, $a_0^U$, ${a'_1}^{L}$, $b_1^L$, $a_1^U$, ${a'_2}^{L}$, $b_2^L$, $a_2^U$, ${a'_3}^{L}$, $a_3^U\}$ need to be verified to estimate the gains of 1-photon and 2-photon states as in Eqs.~(\ref{Eq: Q1L}) and~(\ref{Eq: Q2L}). Then one can numerically choose the optimal values $e_1$ and $e_2$ under constrain $Q_\mu E_\mu \ge Q_1e_1+Q_2e_2$ to lower bound the secure key rate in Eq.~(\ref{Eq:R_SARG1}). This case is defined as {\bf Case-2}. Note that the conditions in Eq.~(\ref{Eq:condition 0}) for case-1 and in Eqs.~(\ref{Eq:condition1}) and~(\ref{Eq:condition2}) for case-2 need to be verified experimentally. In the following, we propose an active and a passive source monitors to estimate these statistical parameters experimentally.

\section{Active and Passive Source Monitors}

\begin{figure}[t]
\begin{center}
\includegraphics[width=0.48\textwidth]{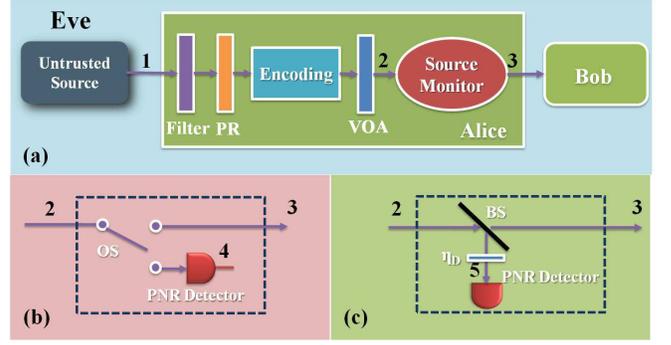}
\end{center}
\caption{(Color online) (a) Schematic diagram of the QKD system with an untrusted source. The untrusted source prepared at P$1$ by Eve, where P$i$ means position $i$ $(i=1,2,3,4,5)$, passes through an optical filter, a phase randomizer (PR), an encoder and a variable optical attenuator
(VOA) with attenuation coefficient $\eta=\eta_s$, $\eta_{d1}$, $\eta_{d2}$, and 0, for the signal, decoy-1, decoy-2, and vacuum source, respectively. Then, the source is sent into a source monitor at P$2$ to estimate the statistical parameters for security analysis, and sent out of Alice's side at P$3$. (b) Schematic diagram of an active source monitor. A high-speed active optical-switch (OS) randomly sends one half of the input optical pulses to a photon-number-resolving (PNR) detector at P$4$ for parameter estimation, and sends the other half to Bob for key generation. (c) Schematic diagram of a passive source monitor. The optical pulses are passively separated into two parts by a beam-splitter (BS) with transmittance $\eta_{BS}$: one goes to a PNR detector with efficiency $\eta_D$ at P$5$, which is modeled by an attenuator with efficiency $\eta_D$ combined with an ideal PNR detector, and the other is sent out of the source monitor.}\label{fig: Source Monitor}
\end{figure}

The schematic diagram of a QKD system with an untrusted source is shown in Fig.~\ref{fig: Source Monitor}(a), where the source is assumed to be completely controlled and prepared by Eve. A source monitor is used to verify the statistical characteristics of the untrusted source in Alice's side. At least two schemes can realize the source monitor: an active scheme shown in Fig.~\ref{fig: Source Monitor}(b)~\cite{UntruQKD_Zhao_08}, and a passive scheme shown in Fig.~\ref{fig: Source Monitor}(c)~\cite{UntruQKD_Peng_08,UntruQKD_Zhao_09,UntruQKD_Peng_09}. Suppose that $P_1(n)$ is the PND of the untrusted source at P$1$ (P$i$ means position $i$ in Fig.~\ref{fig: Source Monitor}), and  $P_3(m,\eta)$ is the PND at P$3$ given that the attenuation coefficient of the VOA is $\eta$. Then one has~\cite{UntruQKD_Peng_08}
\begin{equation}
{P_3}(m,\eta ) = \sum\nolimits_{n = m}^\infty  {{P_1}(n)\left( {\begin{array}{*{20}{c}}
   n  \\
   m  \\
\end{array}} \right)\eta'^m{{(1 - \eta')}^{n- m}}},
\end{equation}
where $\eta '=\eta$ for active scheme and $\eta '=\eta\times\eta_{BS}$ for passive scheme. Due to the definition of $\{a'_m,a_m,b_m\}$,
\begin{equation}\label{Eq:Output PND}
a'_m=P_3(m,\eta_s),\ a_m=P_3(m,\eta_{d1}),\ b_m=P_3(m,\eta_{d2}).
\end{equation}
A full security analysis procedure can be divided in to four steps. {\bf Step1:} Estimate the bounded statistical parameters of the untrusted source based on the experimental measurement results. {\bf Step2:} Verify the conditions shown in Eqs.~(\ref{Eq:condition1}) and~(\ref{Eq:condition2}). {\bf Step3:} Calculate the lower bound of $Q_1$ and $Q_2$ based on Eqs.~(\ref{Eq: Q1L}) and~(\ref{Eq: Q2L}). {\bf Step4:} Estimate the final secure key rate.

\subsection{Active Source Monitor}
In the active source monitor shown in Fig.~\ref{fig: Source Monitor}(b), one half of the optical pulses are randomly sent to a photon-number-resolving (PNR) detector for parameters estimation, and the other half are sent to Bob for key generation~\cite{UntruQKD_Zhao_08}. For simplification, the PNR detector is assumed to be noiseless and the detection efficiency is $1$. Clearly,
\begin{equation}\label{Eq:Active PND Relation}
D(m,\eta)=P_3(m,\eta),\ (m=0,1,2,3,\cdots),
\end{equation}
where $D(m,\eta)$ is the probability that $m$ photoelectrons are recorded by the PNR detector given that the attenuation coefficient of the VOA is $\eta$. Combining the results in Eqs.~(\ref{Eq:Output PND}) and (\ref{Eq:Active PND Relation}), $a'_m=D(m,\eta_s),a_m=D(m,\eta_{d1}),b_m=D(m, \eta_{d2}).$ Clearly, one can bound of parameters $\{a'_m,a_m,b_m\}$ based on the recorded data $D(m,\eta)$. Then one can verify the conditions in Eqs.~(\ref{Eq:condition 0}),~(\ref{Eq:condition1}), and~(\ref{Eq:condition2}), and calculate the final secure key rate.

\subsection{Passive Source Monitor}
As pointed out in \cite{{UntruQKD_Peng_08},UntruQKD_Zhao_09}, it is challenging and inefficient to implement the active scheme. Then, a practical passive scheme is proposed and tested experimentally \cite{UntruQKD_Peng_08}. In the passive source monitor shown in Fig.~\ref{fig: Source Monitor}(c), optical pulses are separated into two parts by a beam splitter (BS) with transmittance $\eta_{BS}$: one goes to a PNR detector with efficiency $\eta_D$, and the other is sent out of Alice's side. For simplification, one set
\begin{equation}\label{Eq:eta_BS and eta_D}
(1-\eta_{BS})\eta_D=\eta_{BS}.
\end{equation}
Then the PND at P$3$ is the same to that at P$5$,
\begin{equation}\label{Eq:Passive PND Relation}
P_5(m,\eta)=P_3(m,\eta).
\end{equation}
If the PNR detector is noiseless, the detected photoelectron distribution $F(m,\eta)$ at P$5$ will be the same to $P_5(m,\eta)$,
\begin{equation}\label{Eq:Passive PED PND Relation}
F(m,\eta)=P_5(m,\eta).
\end{equation}
Based on Eqs. (\ref{Eq:Output PND}), (\ref{Eq:Passive PND Relation}) and (\ref{Eq:Passive PED PND Relation}), one can bound the parameters $\{a'_m, a_m,b_m\}$ with the knowledge of $F(m,\eta)$. In a real system, one needs to consider the practical imperfections of the source monitor. In the following, the effects of statistical fluctuation and detection noise are quantitatively analyzed.

\subsubsection{Infinite Data Size and Noiseless Source Monitor}

Suppose that $M$ is the total number of optical pulses sent from Alice to Bob, while $p'M(=M_s)$, $p_1M(=M_1)$, and $p_2M=(M_2)$ is the number of signal, decoy-1, and decoy-2 pulses, correspondingly.

{\bf Step1.} When the data size $M\to \infty$, one has ${a'_m}^L={a'_m}^U=F(m,\eta_s),\ a_m^L=a_m^U=F(m,\eta_{d1}),\ b_m^L=b_m^U=F(m,\eta_{d2})$.

{\bf Step2.} The conditions in Eqs.~(\ref{Eq:condition1}) and (\ref{Eq:condition2}) turn to
\begin{eqnarray*}
&&\frac{F(k,\eta_s)}{F(k,\eta_{d1})} \ge \frac{F(3,\eta_s)}{F(3,\eta_{d1})} \ge \frac{F(2,\eta_s)}{F(2,\eta_{d1})} \ge \frac{F(1,\eta_s)}{F(1,\eta_{d1})}\ (k\ge 4),\\
&&1+\frac{F(3,\eta_{d1})F(1,\eta_s)-F(1,\eta_{d1})F(3,\eta_s)}{F(2,\eta_{d1})F(3,\eta_s)-F(3,\eta_{d1})F(2,\eta_s)}\frac{F(2,\eta_{d2})}{F(2,\eta_{d1})}>0.
\end{eqnarray*}

{\bf Step3.} In case-1, the gain of 1-photon state is calculated by Eq.~(\ref{Eq: Q1L}) based on the recorded data $F(m,\eta)$, and all the errors are assumed from 1-photon state $e_1=E_\mu Q_\mu/Q_1$. In case-2, the gains of 1-photon and 2-photon states are calculated by Eqs.~(\ref{Eq: Q1L}) and~(\ref{Eq: Q2L}) based on the recorded data  $F(m,\eta)$, and $e_{1(2)}$ are chosen numerically to lower bound the secure key rate.

{\bf Step4.} Calculate the secure key rate for case-1 and case-2 with Eqs.~(\ref{Eq:R 1photon}) and~(\ref{Eq:R_SARG1}), respectively.

\begin{figure}[t]
\begin{center}
\includegraphics[width=0.48\textwidth]{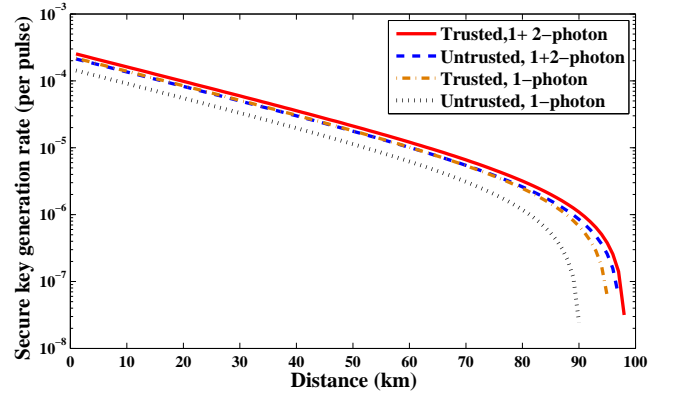}
\end{center}
\caption{(Color online) Simulation results of the SARG04 protocol for the trusted source, compared with the untrusted source in both case-1 and case-2 when the data size is infinite. In the trusted source case, infinite decoy state method is used to estimate the values of ${Q_1,Q_2,e_1,e_2}$ as in Eq.~(\ref{Eq:Trusted source}). The top (red) line is the simulation results for the trusted source, where one considers the contribution from both 1-photon and 2-photon states. The second (yellow) line is the simulation results for the untrusted source in case-2. The third (blue) line is the simulation results for the trusted source, where one considers the contribution from only 1-photon state. The bottom (green) line is the simulation results for the untrusted source in case-1. In all the simulations, the PND for both trusted and untrusted source is assumed to be of Poissonian statistics.}\label{fig: Inf Data}
\end{figure}

For testing the efficiency of the passive scheme, the simulation results for the trusted source are compared with that for the untrusted source (shown in Fig.~\ref{fig: Inf Data}), while the data size is infinite. The PND for the trusted and the untrusted source is assumed to be of Poissonian
statistics to perform simulations. The error correction efficiency $f(E_\mu)=1.22$. The transmittance $\eta_{BS}$ of the BS is 0.13 and the detection efficiency $\eta_D$ of the PNR detector is 0.15. The other experimental parameters are cited from the GYS experiment \cite{GYS_04} as shown in Table~\ref{tab:para}, where $\eta_{Bob}$ is the efficiency of Bob's detection, $e_0$ is the probability that a dark count hit the erroneous detector in Bob's side. Suppose the average photon number (APN) for signal, decoy-1 and decoy-2 sources are $\mu$, $v_1$ and $v_2$, respectively. The conditions in Eqs.~(\ref{Eq:condition1}) and~(\ref{Eq:condition2}) turn to $\frac{{{e^{ - \mu }}{\mu ^k}}}{{{e^{ - {v_1}}}{v_1}^k}} \ge \frac{{{e^{ - \mu }}{\mu ^3}}}{{{e^{ - {v_1}}}{v_1}^3}} \ge \frac{{{e^{ - \mu }}{\mu ^2}}}{{{e^{ - {v_1}}}{v_1}^2}} \ge \frac{{{e^{ - \mu }}\mu }}{{{e^{ - {v_1}}}{v_1}}}$ for all $k\ge4,$ and $1-\frac{v_2}{v_1}\frac{v_1+\mu}{\mu}>0$. As shown in Fig.~\ref{fig: Inf Data}, the performance of the untrusted source based on the passive source monitor is very close to that of the trusted source, and the 2-photon state makes positive contribution to the secure key rate.

\begin{table}[htbp]
\caption{The simulation parameters for Figs.~\ref{fig: Inf Data}-\ref{fig: Detection Noise}.}
\begin{ruledtabular}
\begin{tabular}{ccccccc}
$\eta_D$ &$\eta_{BS}$&$\eta_{Bob}$&$\alpha$&$Y_0$&$e_{\det}$&$e_0$\\
\hline $0.15$& 0.13&0.045&0.21&$1.7 \times 10^{-6}$&3.3\%&0.5
\end{tabular}
\end{ruledtabular}
\label{tab:para}
\end{table}

\subsubsection{Finite Data Size and Noiseless Source Monitor}

\begin{figure}[htbp]
\begin{center}
\includegraphics[width=0.48\textwidth]{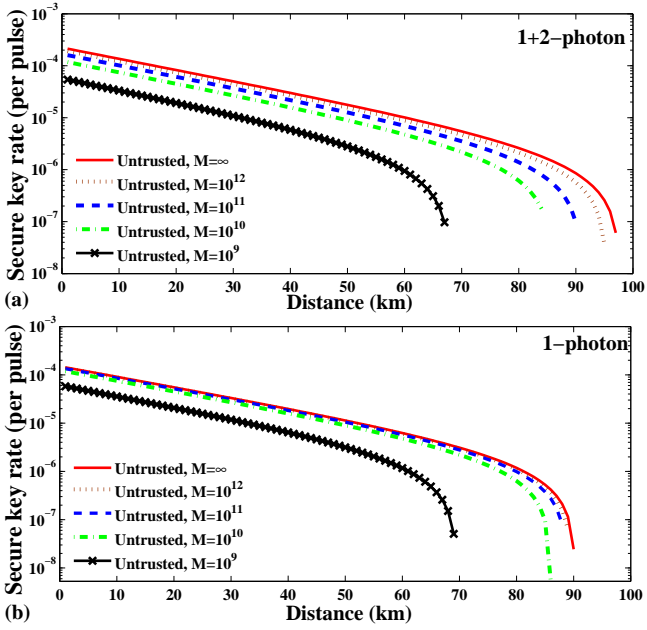}
\end{center}
\caption{(Color online) Simulation results for the SARG04 protocol with an untrusted source in case-1 and case-2 with data size $M$, based on the passive source monitor. The PND of the untrusted source is assumed to be Poissonian. The other experimental parameters are cited from Table \ref{tab:para}. (a)~Simulation results in case-2 with data size $M ={\infty},{10^{12}},{10^{11}},{10^{10}},{10^9}$, respectively. (b)~Simulation results in case-1 with data size $M ={\infty},{10^{12}},{10^{11}},{10^{10}},{10^9}$, respectively. The confidence level is set to be $1-10^{-6}$.}\label{fig: Fin Data}
\end{figure}

\begin{figure}[htbp]
\begin{center}
\includegraphics[width=0.48\textwidth]{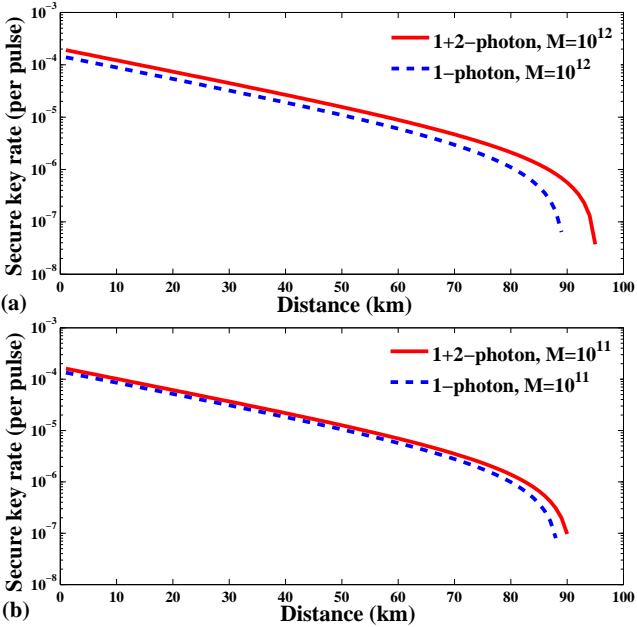}
\end{center}
\caption{(Color online) Comparison between case-1 and case-2 for the SARG04 protocol with an untrusted source: (a)~with data size $M=10^{12}$; (b)~with data size $M=10^{11}$. The PND of the untrusted source is assumed to be Poissonian. The other experimental parameters are cited from Table~\ref{tab:para}. The confidence level is set to be $1-10^{-6}$.}\label{fig: Fin Data Compare}
\end{figure}

Suppose that the data size $M$ is finite. Let $j_m^{s}$, $j_m^{d1}$, and $j_m^{d2}$ denote the number of detected signal, decoy-1 and decoy-2 pulses at P$5$ given the PNR detector records $m$ photoelectrons.

{\bf Step1.} The APN of signal source is $\mu\sim O(10^{-1})$ while the APN of decoy-1 source is $v_1\sim O(10^{-2})$. When the data size is finite (e.g. $M=10^{12}$), one may always observe that $j_m^{d1}=0$ for all $m>J$ while $j_J^{d1}>0$ (e.g. $J=10$) in a real experiment, which infers that the counts caused by the photon number states $m>J$ in decoy-1 source can be ignored in the experiment. For instance, given that the PND of the decoy-1 source is Poissonian with an APN $v_1=0.01$, the probability that the decoy-1 source sends out photon number states $n>10$ is less than $10^{-25}$, which can be ignored for data size $M=10^{12}$. To estimate $Q_1$ and $Q_2$, and verify the conditions in Eqs.~(\ref{Eq:condition 0}), (\ref{Eq:condition1}), and (\ref{Eq:condition2}), one only needs to bound the parameters $\{{a'_m}^L,a_m^U,b_n^L\}$ for $m=0,1,\cdots,J$ and $n=0,1,2$ with finite data size (see Appendix B). Using the \emph{random sampling theory} \cite{RandomSampling_37}, each $ F(m,\eta_s) \in [j_m^s /M_s- \varepsilon',j_m^s /M_s + \varepsilon' ]$ with a confidence level $1-2\exp(-M_s {\varepsilon'}^2/2)$ for signal pulses, and each $ F(m,\eta_{1(2)}) \in [j_m^{d1(2)} /M_{1(2)} - \varepsilon_{1(2)} ,j_m^{d1(2)} /M_{1(2)} + \varepsilon_{1(2)} ]$ with a confidence level $1-2\exp(-M_1(2)\varepsilon_{1(2)}^2/2)$ for decoy-1(2) pulses can be estimated. Simultaneously, $ F(m,\eta_s) \in [j_m^s /M_s- \varepsilon',j_m^s /M_s + \varepsilon' ]$, $ F(m,\eta_1) \in [j_m^{d1} /M_{1}- \varepsilon_1,j_m^{d1} /M_{1} + \varepsilon_1]$ for $m=0,1,2,\cdots,J$, and $ F(n,\eta_2) \in [j_n^{d2} /M_{2}- \varepsilon_2 ,j_n^{d2} /M_{2} + \varepsilon_2]$ for $n=0,1,2$ are approximately estimated with a confidence level $\alpha=1-2(J+1) \exp(-M_s{\varepsilon'}^2/2)-2(J+1)\exp(-M_1 {\varepsilon_1}^2/2)-6\exp(-M_2{\varepsilon_2}^2/2)$. From Eqs.~(\ref{Eq:Output PND}) and (\ref{Eq:Passive PED PND Relation}), one gets
\begin{equation*}
{a'_m}^L=\frac{k_m^s}{M_s}-\varepsilon',\ a_m^U=\frac{k_m^{d1}}{M_1}+\varepsilon_1,\ b_n^L=\frac{k_n^{d2}}{M_2}-\varepsilon_2,
\end{equation*}
for $m=0,1,2,\cdots,J$ and $n=0,1,2$ with confidence level $\alpha$.

{\bf Step2.} It is challenging to verify directly the condition in Eq.~(\ref{Eq:condition1}) with finite data size:~a) In hardware, the PNR detector is required to discriminate the photon number $n=0,1,\cdots,\infty$;~b) When the photoelectron number $m$ is large enough, one always gets $k_m^{s}= k_m^{d1}=0$. One needs a reasonable cutoff value of $m$. Suppose $j_m^{d1}=0$ for all $m>J$ while $j_{J}^{d1}>0$. To lower bound the gains of $Q_1$ and $Q_2$ with finite data size, one can replace the condition in Eq.~(\ref{Eq:condition1}) as
\begin{equation}\label{Eq:condition_Finite_data}
\frac{{{a'_k}^{L}}}{{a_k^{U}}} \ge \frac{{{a'_3}^{L}}}{{a_3^{U}}} \ge \frac{{{a'_2}^{L}}}{{a_2^{U}}} \ge \frac{{{a'_1}^{L}}}{{a_1^{U}}},\ (\rm{for\ all}\ 4\le k \le J),
\end{equation}
where the PNR detector is only required to discriminate photon number $n=0,1,\cdots,J$ (see Appendix B for details). The condition in Eq.~(\ref{Eq:condition2}) turns to
\begin{equation}
1+\frac{F(3,\eta_{d1})F(1,\eta_s)-F(1,\eta_{d1})F(3,\eta_s)}{F(2,\eta_{d1})F(3,\eta_s)-F(3,\eta_{d1})F(2,\eta_s)}\frac{F(2,\eta_{d2})}{F(2,\eta_{d1})}>0.
\end{equation}

{\bf Step3.} If the conditions in step2 are satisfied, one can lower bound the parameters $Q_1$ and $Q_2$.

{\bf Step4.} Calculate the secure key rate for case-1 and case-2 with Eqs.~(\ref{Eq:R 1photon}) and~(\ref{Eq:R_SARG1}), respectively.

For testing the effects of finite data size, we choose an untrusted source of Poissonian statistics to perform simulations in both case-1 and case-2. The error correction efficiency $f(E_\mu)$ are chosen to be 1.22. The other experimental parameters are cited from Table I. Simulation results for case-2 and case-1 are shown in Fig.~\ref{fig: Fin Data}(a) and (b), and the data size are set to be $M=\infty$, $10^{12}$, $10^{11}$, $10^{10}$ and $10^{9}$, respectively. To compare the two cases more clearly, Fig.~\ref{fig: Fin Data Compare} shows the simulation results for case-1 and case-2 with $M=10^{12}$ and $10^{11}$, respectively. In all the above simulations, the confidence level is set to be $\alpha=1-10^{-6}$. The simulation results show that statistical fluctuation has negative effect on performance of the QKD system. When the data size is large enough, the 2-photon state has positive contribution to the secure key rate.

\subsubsection{Finite Data Size and Source Monitor with Random Additive Detection Noise}
Given a PNR detector with an independent additive detection noise $y$, the detected photoelectron number $m'$, and the photon number $m$ at P$5$ satisfy $m'=m+y$. One can calculate the lower and upper bound of PND $P_5(m,\eta)$ at P$5$ based on the photoelectron distribution $F(m,\eta)$
with a high confidence level, given that the distribution of the detection noise $N(y)$ is known by Alice.

The dark count is the main kind of detection noise for the PNR detector such as time multiplexing detector (TMD) \cite{PNR_TMD_03,PNR_TMD_04}, transition-edge sensor (TES) \cite{PNR_TES_05}, or a threshold detector together with a VOA \cite{PNR_OnOff_05}. In case of independent Poisson statistics noise, the probability of detecting $m'$ photoelectrons is $F(m',\eta)=\sum_{d= 0}^{m'} {N(y=m'-d)P_5(d,\eta)}$ where $N(y=d)= e^{ - \lambda }{\lambda ^d }/{d!}$ is the probability that $d$ dark counts occur in the PNR detector, and $\lambda$ is the average dark-count rate. Then, one has
\begin{equation*}\label{Eq:PND and PED and Noise}
\left[ {\begin{array}{*{20}{c}}
{{P_5}(0,\eta )} \\
{{P_5}(1,\eta )} \\
{{P_5}(2,\eta )} \\
{{P_5}(3,\eta )} \\
\end{array}} \right] = \left[ {\begin{array}{*{20}{c}}
{F(0,\eta )} & 0 & 0 & 0 \\
{F(1,\eta )} & {F(0,\eta )} & 0 & 0 \\
{F(2,\eta )} & {F(1,\eta )} & {F(0,\eta )} & 0 \\
{F(3,\eta )} & {F(2,\eta )} & {F(1,\eta )} & {F(0,\eta )} \\
\end{array}} \right]\left[ {\begin{array}{*{20}{c}}
{{e^\lambda }} \\
{ - \lambda {e^\lambda }} \\
{{\lambda ^2}{e^\lambda }/2} \\
{ - {\lambda ^3}{e^\lambda }/6} \\
\end{array}} \right].
\end{equation*}

{\bf Step1.} Using \emph{random sampling theory} \cite{RandomSampling_37}, simultaneously, $ F(m,\eta_s) \in [j_m^s /M_s- \varepsilon',j_m^s /M_s + \varepsilon' ]$, $ F(m,\eta_{d1}) \in [j_m^{d1} /M_{1}- \varepsilon_{1} ,j_m^{d1} /M_{1} + \varepsilon_{1} ]$, and $ F(m,\eta_{d2}) \in [j_n^{d2} /M_{2}- \varepsilon_{2} ,j_n^{d2} /M_{2} + \varepsilon_{2}]$ for $m=0,1,2,\cdots,J$ and $n=0,1,2$ are estimated with a confidence level $1-2(J+1)\exp(-M_s {\varepsilon'}^2/2)-2(J+1)\exp(-M_1 {\varepsilon_1}^2/2)-6\exp(-M_2{\varepsilon_2}^2/2)$.
Then, one yields
{\setlength\arraycolsep{1pt}\begin{eqnarray*}
{a_0} &\le& {e^\lambda }(\frac{{j_{m = 0}^{d1}}}{{{M_1}}} + \varepsilon '),\\
{b_0} &\ge& {e^\lambda }(\frac{{j_{m = 0}^{d2}}}{{{M_2}}} - {\varepsilon _2}),\\
a{'_0} &\ge& {e^\lambda }(\frac{{j_{m = 0}^s}}{{{M_s}}} - \varepsilon '), \\
 {a_1} &\le& {e^\lambda }(\frac{{j_{m = 1}^{d1}}}{{{M_1}}} + {\varepsilon _1}) - \lambda {e^\lambda }(\frac{{j_{m = 0}^{d1}}}{{{M_1}}} - {\varepsilon _1}),\\
 {b_1} &\ge& {e^\lambda }(\frac{{j_{m = 1}^{d2}}}{{{M_2}}} - {\varepsilon _2}) - \lambda {e^\lambda }(\frac{{j_{m = 0}^{d2}}}{{{M_2}}} + {\varepsilon _2}),\\
a{'_1} &\ge& {e^\lambda }(\frac{{j_{m = 1}^s}}{{{M_s}}} - \varepsilon ') - \lambda {e^\lambda }(\frac{{j_{m = 0}^s}}{{{M_s}}} + \varepsilon '), \\
 {a_2} &\le& {e^\lambda }(\frac{{j_{m = 2}^{d1}}}{{{M_1}}} + {\varepsilon _1}) - \lambda {e^\lambda }(\frac{{j_{m = 1}^{d1}}}{{{M_1}}} - {\varepsilon _1}) + \frac{{{\lambda ^2}}}{2}{e^\lambda }(\frac{{j_{m = 0}^{d1}}}{{{M_1}}} + {\varepsilon _1}),\\
 {b_2} &\ge& {e^\lambda }(\frac{{j_{m = 2}^{d2}}}{{{M_2}}} - {\varepsilon _2}) - \lambda {e^\lambda }(\frac{{j_{m = 1}^{d2}}}{{{M_2}}} + {\varepsilon _2}) + \frac{{{\lambda ^2}}}{2}{e^\lambda }(\frac{{j_{m = 0}^{d2}}}{{{M_2}}} - {\varepsilon _2}),\\
 a{'_2} &\ge& {e^\lambda }(\frac{{j_{m = 2}^s}}{{{M_s}}} - \varepsilon ') - \lambda {e^\lambda }(\frac{{j_{m = 1}^s}}{{{M_s}}} + \varepsilon ') + \frac{{{\lambda ^2}}}{2}{e^\lambda }(\frac{{j_{m = 0}^s}}{{{M_s}}} - \varepsilon '),\\
 {a_3} &\le&{e^\lambda }(\frac{{j_{m = 3}^{d1}}}{{{M_1}}} + {\varepsilon _1}) - \lambda {e^\lambda }(\frac{{j_{m = 2}^{d1}}}{{{M_1}}} - {\varepsilon _1})+ \frac{{{\lambda ^2}}}{2}{e^\lambda }(\frac{{j_{m = 1}^{d1}}}{{{M_1}}} + {\varepsilon _1})\\
&& - \frac{{{\lambda ^2}}}{2}{e^\lambda }(\frac{{j_{m = 0}^{d1}}}{{{M_1}}} - {\varepsilon _1}),\\
a{'_3} &\ge&{e^\lambda }(\frac{{j_{m = 3}^s}}{{{M_s}}} - \varepsilon ') - \lambda {e^\lambda }(\frac{{j_{m = 2}^s}}{{{M_s}}} + \varepsilon ')
+\frac{{{\lambda ^2}}}{2}{e^\lambda }(\frac{{j_{m = 1}^s}}{{{M_s}}} - \varepsilon ')\\
&& - \frac{{{\lambda ^2}}}{2}{e^\lambda }(\frac{{j_{m = 0}^s}}{{{M_s}}} + \varepsilon ').
\end{eqnarray*}}Our analysis is not limited to the Poissonian noise case. Generally, when the random-positive detection noise y with distribution $N(y)$ is known to Alice, one can still use the same method in~\cite{Untrusted_Xu_10} to estimate the parameters $\{$${a'_0}^L$, $b_{0}^L$, $a_0^U$, ${a'_1}^{L}$, $b_1^L$, $a_1^U$, ${a'_2}^{L}$, $b_2^L$, $a_2^U$, ${a'_3}^{L}$, $a_3^U\}$ with a certain confidence level.

{\bf Step2.} Using the same method, one can estimate the bound values $\{{a'_k}^{L},a_k^U\}$ for $4\le k\le J$, and verify the conditions in Eqs.~(\ref{Eq:condition2}) and~(\ref{Eq:condition_Finite_data}). Since the expressions of $\{{a'_k}^{L},a_k^U\}$ for $4\le k\le J$ are much complex and trivial, we assume the above conditions are satisfied as in \cite{UntruQKD_Peng_08,{UntruQKD_Wang_09}}.

{\bf Step3.} Lower bound the parameters $Q_1$ and $Q_2$.

{\bf Step4.} Calculate the secure key rate for case-1 and case-2 with Eqs.~(\ref{Eq:R 1photon}) and~(\ref{Eq:R_SARG1}), respectively.

\begin{figure}[t]
\begin{center}
\includegraphics[width=0.48\textwidth]{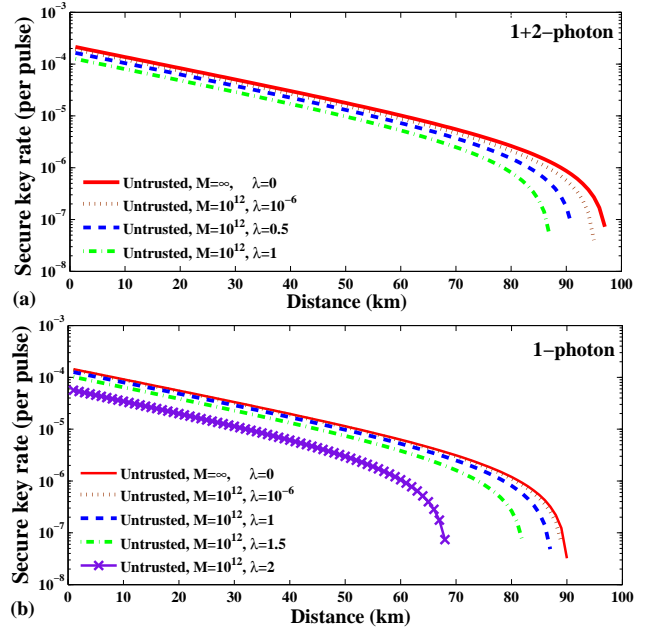}
\end{center}
\caption{(Color online) Simulation results for the SARG04 protocol with an untrusted source based on the passive scheme: (a)~with finite data size $M=10^{12}$, and the average dark count rate of the Poissonian detection noise $\lambda=0,10^{-6},0.5$, and $1$, respectively,
in case-2; (b)~with finite data size $M=10^{12}$, and the average dark count rate of the Poissonian detection noise $\lambda=0,10^{-6},1,1.5$, and $2$, respectively, in case-1. The PND of the untrusted source and the distribution of detection noise are assumed to be Poissonian. The other experimental parameters are cited from Table~\ref{tab:para}. The confidence level is $1-10^{-6}$.}\label{fig: Detection Noise}
\end{figure}

For testing the effect of dark count noise, the simulation results for case-2 are shown in Fig.~\ref{fig: Detection Noise}(a), with finite data size  $M=10^{12}$ and average dark count rate $\lambda =0$, $10^{-6}$, $0.5$, and $1$, respectively. The simulation results for case-1 are shown in Fig.~\ref{fig: Detection Noise}(b) with $M=10^{12}$ and $\lambda=0$, $10^{-6}$, $1$, $1.5$ and $2$, respectively. The confidence level is set to be $\alpha=1-10^{-6}$.

\section{Summary and Conclusion Remark}
In summary, we have shown the unconditional security of the SARG04 protocol with an untrusted source, given that the bound of some key statistical parameters of the untrusted source are known by Alice and Bob. The analytical expression for the lower bound of the gain of 2-photon state is derived. Furthermore, an active and a passive source monitors are proposed to verify these parameters experimentally. Finally, we analyze the effects of the practical imperfections in the passive source monitor quantitatively, such as finite data size and additive detection noise. Asymptotically, the performance of the QKD system with an untrusted source combined with passive source monitor is very close to that of a trusted source. Our results can be directly applied to plug-and-play QKD system with SARG04 protocol.

\begin{acknowledgments}
This work is supported by the Key Project of National Natural Science Foundation of China (Grant No. 60837004) and National Hi-Tech Research and Development (863) Program.
\end{acknowledgments}

\appendix

\section{Lower bound of $Q_1$ and $Q_2$ for SARG04 Protocol with an untrusted source}

Suppose Alice sends $M$ pulses to Bob in the whole quantum process. At any time $i$, where $i\in{\{1,2,\cdots,M\}}$, Alice randomly produces vacuum ($\left| 0 \right\rangle\left\langle 0 \right|$), decoy-1 ($\rho_{1i}=\sum_{n = 0}^\infty {a_{ni} \left| n \right\rangle }\left\langle n \right|$),
decoy-2 ($\rho_{2i}=\sum_{n = 0}^\infty {b_{ni} \left| n \right\rangle }\left\langle n \right|$), and signal ($\rho _{si} = \sum_{n = 0}^\infty {a'_{ni} \left| n \right\rangle } \left\langle n \right|$) source with the probability $p_0$, $p_1$, $p_2$, and $p'$, respectively, where $\{a_{ni},\ b_{ni},\ a'_{ni}\}$ are controlled and prepared by Eve. Following the methods in \cite{UntruQKD_Wang_08,{UntruQKD_Wang_09}}, some definitions are necessary for further analysis. \emph{Definition 1.} In the protocol, Alice sends $M$ pulses, and Bob gets $M$ observations. If Bob's detector click at time $i$, we say that ``the $i$th pulse from Alice has caused a count''.
\emph{Definition 2.} Sets $C$ and $c_n$: Set $C$ contains any pulse that has caused a count; set $c_n$ contains any $n$-photon pulse that has caused a count. \emph{Definition 3.} Denote the lower (upper) bound of $\{a'_{ni},\ a_{ni},\ b_{ni}\}$ as $\{{a'_n}^{L(U)},\ a_n^{L(U)},\ b_n^{L(U)}\}$ for $i=1,2,\cdots,M$.

Define ${d_{ki}} = \frac{1}{{p_1{a_{ki}}+p_2{b_{ki}} + {p{'}}a'_{ki}}}$ and ${D_k} = \sum_{i \in {c_k}} {{d_{ki}}}.
$ If the $i$th pulse contains zero photon, the probability that it comes from the vacuum source is ${P_{vi|0}} = p_0d_{0i}$. Therefore, the number of counts caused by vacuum source is $N_0 =\sum_{i \in {c_0}}{p_{vi|0}}=\sum_{i \in {c_0}}{{p_0}{d_{0i}}}.$ Similarly, if the $i$th pulse contains zero photon, the probability that it comes from the decoy-1, decoy-2 and signal source are $P_{1i|0}=p_1a_{0i}d_{0i}$, $P_{2i|0}=p_2b_{0i} d_{0i}$, and $P_{si|0}=p'a'_{0i}d_{0i}$, respectively. Then the number of counts caused by zero photon state in decoy-1, decoy-2 and signal source are ${n_{0d1}}= \sum_{i \in {c_0}} {p_1{a_{0i}}} {d_{0i}},$ ${n_{0d2}}= \sum_{i \in {c_0}} {p_2{b_{0i}}} {d_{0i}},$ and ${n_{0s}} =\sum_{i \in {c_0}} {p{'}{a'_{0i}}} {d_{0i}},$ respectively. It is clear that
\begin{eqnarray}\label{Eq:n0}
\nonumber n_{0d1}^{U} = \frac{{p_1a_0^{U}{N_0}}}{{{p_0}}} \ge &{n_{0d1}}& \ge \frac{{p_1a_0^{L}{N_0}}}{{{p_0}}} = n_{0d1}^{L},\\
n_{0d2}^{U} = \frac{{p_2b_0^{U}{N_0}}}{{{p_0}}} \ge &{n_{0d2}}& \ge \frac{{p_2b_0^{L}{N_0}}}{{{p_0}}} = n_{0d2}^{L},\\
\nonumber n_{0s}^{U} = \frac{{{p{'}}{a'_0}^{U}{N_0}}}{{{p_0}}} \ge &{n}_{0s}& \ge \frac{{{p{'}}{a'_0}^{L}{N_0}}}{{{p_0}}} = n_{0s}^{L}.
\end{eqnarray} The main result in \cite{UntruQKD_Wang_08} is to derive the lower bound of $D_1$ based on 3-intensity decoy state method (which can be seen as a special case of the 4-intensity decoy state method $p_2=0$ in this paper),
\begin{equation}\label{Eq: D1L}
D_1\ge{D_1^L}= \frac{{\frac{{{a'_2}^{L}}}{p_1}{N_{d1}} - \frac{{a_2^U}}{p^{'}}{N_S} - \frac{{{a'_2}^{L}}}{p_1}{n_{0d1}^U}+ \frac{{a_2^{U}}}{{{p{'}}}}n_{0s}^L}}{{{a'_2}^{L}a_1^{U} - a_2^{U}{a'_1}^{L}}},
\end{equation} under condition
$\frac{{{a'_k}^{L}}}{{a_k^{U}}} \ge \frac{{{a'_2}^{L}}}{{a_2^{U}}} \ge \frac{{{a'_1}^{L}}}{{a_1^{U}}}$ for all $k \ge 3$. Further, one can lower bound the gain of 1-photon state in signal source, $Q_1={p{'}}\sum\limits_{i \in {c_1}} {a'_{1i}{d_{1i}}}\frac{1}{p'M}\ge \frac{{a'_1}^{L}{D_1^L}}{M}$ as shown in Eq.~(\ref{Eq: Q1L}). In the following, the lower bound of $Q_2$ is derived. The number of counts caused by decoy-1(2) and signal sources are
\begin{eqnarray}\label{Eq: NsNd}
\nonumber {N_{d1}} &=& {n_{0d1}} + p_1\sum\limits_{i \in {c_1}} {{a_{1i}}{d_{1i}}}  + p_1\sum\limits_{i \in {c_2}} {{a_{2i}}{d_{2i}}}  + p_1\sum\limits_{k = 3}^\infty {\sum\limits_{i \in {c_k}} {{a_{ki}}{d_{ki}}} },\\
\nonumber {N_{d2}} &=& {n_{0d2}} + p_2\sum\limits_{i \in {c_1}} {{b_{1i}}{d_{1i}}}  + p_2\sum\limits_{i \in {c_2}} {{b_{2i}}{d_{2i}}}  + p_2\sum\limits_{k = 3}^\infty {\sum\limits_{i \in {c_k}} {{b_{ki}}{d_{ki}}} },\\
\nonumber {N_s} &=& n_{0s} + {p{'}}\sum\limits_{i \in {c_1}} {a'_{1i}{d_{1i}}}  + {p{'}}\sum\limits_{i \in {c_2}} {a'_{2i}{d_{2i}}}  + {p{'}}\sum\limits_{k = 3}^\infty {\sum\limits_{i \in {c_k}} {a'_{ki}{d_{ki}}} },\\
\end{eqnarray}
which can be rewritten as
\begin{eqnarray}
\label{Eq:Nd1} &{N_{d1}}&= {n_{0d1}} + p_1a_1^{U}{D_1} + p_1a_2^{U}{D_2} + p_1\Lambda_1 - {\xi _1},\\
\label{Eq:Nd2}&{N_{d2}}&= {n_{0d2}} + p_2b_1^{L}{D_1} + p_2b_2^{L}{D_2} + p_2\Lambda_2 + {\xi _2},\\
\label{Eq:Nss}&{N_s}&=n_{0s} + {p{'}}{a'_1}^{L}{D_1} + {p{'}}{a'_2}^{L}{D_2} + {p{'}}{\Lambda {'}} + {\xi _3},\\
\nonumber
\end{eqnarray}
where $\Lambda_1 = \sum_{k = 3}^{\infty} {a_k^{U}}D_k,\ {\Lambda _2} = \sum_{k = 3}^{\infty} {b_k^L} D_k,\ {\Lambda {'}} = \sum_{k = 3}^{\infty} {{a'_k}^{L}} D_k,$ and ${\xi _1}\ge0,\ {\xi _2}\ge0,\ {\xi _3}\ge0.$ Define
\begin{equation*}{\xi} = {\Lambda {'}} - {{{a'_3}^{L}}}/{{a_3^{U}}}\Lambda_1,\end{equation*}
and assume $\frac{{{a'_k}^{L}}}{{a_k^{U}}} \ge \frac{{{a'_3}^{L}}}{{a_3^{U}}} \ge \frac{{{a'_2}^{L}}}{{a_2^{U}}} \ge \frac{{{a'_1}^{L}}}{{a_1^{U}}}\ $ for all $k \ge 4$,
which leads to ${\xi } \ge 0$, one has
\begin{equation}\label{Eq:Ns}
{N_s}=n_{0s} + {p{'}}{a'_1}^{L}{D_1} + {p{'}}{a'_2}^{L}{D_2} + {p{'}}{\frac{{a'_3}^{L}}{{a_3}^{U}}}\Lambda_1 +p'\xi+ {\xi _3}.
\end{equation}
Combining Eqs.~(\ref{Eq:Nd1}) and~(\ref{Eq:Ns}), one has
\begin{widetext}
\begin{equation*}
D_2 = \frac{{\frac{{{a'_3}^{L}}}{p_1}{N_{d1}} - \frac{{a_3^{U}}}{p{'}}{N_s} - \frac{{{a'_3}^{L}}}{p_1}{n_{0d1}} + \frac{{a_3^{U}}}{{{p{'}}}}n_{0s} + (a_3^{U}{a'_1}^{L} - {a'_3}^{L}a_1^{U}){D_1} + \frac{{{a'_3}^{L}}}{p_1}{\xi _1} + \frac{a_3^{U}}{p{'}}({\xi _3} + p'{\xi})}}{{{a'_3}^{L}a_2^{U} - a_3^{U}{a'_2}^{L}}}.
\end{equation*}
\end{widetext}
Since ${\xi _1}$, ${\xi _3}$ and ${\xi}$ are all non-negative, $ {{a'_3}^{L}a_2^{U}- a_3^{U}{a'_2}^{L}}\ge 0$ and $ {{a'_3}^{L}a_1^{U} - a_3^{U}{a'_1}^{L}}\ge 0$, one has
\begin{equation}\label{Eq:D2}
{D_2} \ge \frac{{\frac{{{a'_3}^{L}}}{p_1}{N_{d1}} - \frac{{a_3^{U}}}{{{p{'}}}}{N_S} - \frac{{{a'_3}^{L}}}{p_1}{n_{0d1}^U} + \frac{{a_3^{U}}}{{{p{'}}}}n_{0s}^L + (a_3^{U}{a'_1}^{L} - {a'_3}^{L}a_1^{U}){{{D_1^U}}}}}{{{a'_3}^{L}a_2^{U} - a_3^{U}{a'_2}^{L}}}.
\end{equation}
It is clear that $N_{d2} \ge n_{0d2}^L+{p_2}{b_1^L}{D_1}+{p_2}{b_2^L}{D_2}$. Then one has
\begin{equation}\label{Eq:D1U}
D_1 \le \frac{N_{d2}-n_{0d2}^L}{p_2b_1^L}-\frac{b_2^L}{b_1^L}D_2={D_1^U}.
\end{equation}
Combine the results of Eqs.~(\ref{Eq:D2}) and~(\ref{Eq:D1U}), one has
\begin{equation}\label{Eq:D2L}
{D_2^L}=\frac{{\frac{{{a'_3}^{L}}}{p_1}{N_{d1}} - \frac{{a_3^{U}}}{{{p{'}}}}{N_S} - \frac{{{a'_3}^{L}}}{p_1}{n_{0d1}^U} + \frac{{a_3^{U}}}{{{p{'}}}}n_{0s}^L + (a_3^{U}{a'_1}^{L} - {a'_3}^{L}a_1^{U}){\frac{N_{d2}-n_{0d2}^L}{p_2b_1^L}}}}{c({{a'_3}^{L}a_2^{U} - a_3^{U}{a'_2}^{L}})},
\end{equation}
under conditions  $\frac{{{a'_k}^{L}}}{{a_k^{U}}} \ge \frac{{{a'_3}^{L}}}{{a_3^{U}}} \ge \frac{{{a'_2}^{L}}}{{a_2^{U}}} \ge \frac{{{a'_1}^{L}}}{{a_1^{U}}}\ $ for all $k \ge 4$ and  $c=1+\frac{a_3^{U}{a'_1}^{L} - {a'_3}^{L}a_1^{U}}{{a'_3}^{L}a_2^{U} - a_3^{U}{a'_2}^{L}}\frac{b_2^L}{b_1^L}> 0$. Further, one can estimate the lower bound of gain of 2-photon state in signal source $Q_2\ge\frac{{a'_2}^L{D_2^L}}{M}$ as shown in Eq.~(\ref{Eq: Q2L}).

\section{Verify the condition in Equation (\ref{Eq:condition1}) in finite data size}

Suppose that one observe $j_m^{d1}=0$ for all $m>J$ while $j_J^{d1}>0$, and $j_m^{s}=0$ for all $m>J'$ while $j_{J'}^s>0$ $(J'\ge J)$ in a real experiment. Similar to Eqs.~(\ref{Eq: NsNd}),
\begin{equation*}
Q_{d1}=\sum_{k=0}^{J}Q_k^{d1}+\sum_{k=J+1}^{\infty}Q_k^{d1},
\end{equation*}
where $Q_{d1}=N_{d1}/M_1$ is the count rates of the decoy-1 source, and $Q_k^{d1}=p_1\sum_{i \in {c_k}} {{a_{ki}}{d_{ki}}}/M_1 $ is the gain of $k$-photon state in decoy-1 source, which can be explained as the probability that Alice produces a $k$-photon pulse in decoy-1 source and the pulse causes a count at Bob's detectors. Clearly, $Q_k^{d1}\le a_k$ and
$Q_{d1}\le\sum_{k=0}^{J}Q_k^{d1}+\sum_{k=J+1}^{\infty}a_k,$ which infers,
\begin{equation*}
\nonumber {N_{d1}} \le {n_{0d1}} + p_1\sum\limits_{k = 1}^J {\sum\limits_{i \in {c_k}} {{a_{ki}}{d_{ki}}} }+M_1P_J,
\end{equation*}
where $P_J=\sum_{k=J+1}^{\infty}a_k$. Using the {\emph Clopper-Pearson confidence interval theory} \cite{CP}, one can upper bound
$P_J$ with a confidence level $1-\alpha$, where $(1-P_J^U)^{M_1}=\alpha/2$ and $P_J^U\sim {\frac{1}{M_1}}$
is the upper bound of $P_J$. Similar to Eqs.~(\ref{Eq: NsNd}), one has
\begin{eqnarray*}
{N'_{d1}} &\le& {n_{0d1}} + p_1\sum\limits_{i \in {c_1}} {{a_{1i}}{d_{1i}}}  + p_1\sum\limits_{i \in {c_2}} {{a_{2i}}{d_{2i}}}  + p_1\sum\limits_{k = 3}^J {\sum\limits_{i \in {c_k}} {{a_{ki}}{d_{ki}}} },\\
{N_s}&\ge& n_{0s} + {p{'}}\sum\limits_{i \in {c_1}} {a'_{1i}{d_{1i}}}  + {p{'}}\sum\limits_{i \in {c_2}} {a'_{2i}{d_{2i}}}  + {p{'}}\sum\limits_{k = 3}^J {\sum\limits_{i \in {c_k}} {a'_{ki}{d_{ki}}} },
\end{eqnarray*}
where $N'_{d1}=N_{d1}-M_1P_J^U$. Then one can calculate the lower bounds of $D_1$ and $D_2$ can be expressed the same as Eqs.~(\ref{Eq: D1L})
and~(\ref{Eq:D2L}) except replacing the $N_{d1}$ by $N'_{d1}$, and
the condition in Eq.~(\ref{Eq:condition1}) can be replaced by
\begin{equation*}
\frac{{{a'_k}^{L}}}{{a_k^{U}}} \ge \frac{{{a'_3}^{L}}}{{a_3^{U}}} \ge \frac{{{a'_2}^{L}}}{{a_2^{U}}} \ge \frac{{{a'_1}^{L}}}{{a_1^{U}}},\ (\rm{for\ all}\ 4\le k \le J).
\end{equation*}

\end{document}